\documentstyle[12pt]{article}
\begin{document}

%New Commands
\newcommand{\siml}{\stackrel{<}{\sim}}
\newcommand{\simg}{\stackrel{>}{\sim}}
%\baselineskip=1.2\baselineskip
%\baselineskip=1.8\baselineskip

%single vs. double space 
%\renewcommand{\baselinestretch}{2}

\noindent
\begin{center}
{\large\bf
Graded persisting activity of heterogeneous \\
neuron ensembles subject to white noises
%Robust activity of ensemble neurons@\\
%with graded persisting firings
%\footnote{E-print: cond-mat/0401072}
\footnote{E-print: q-bio.NC/0401009
(http://xxx.yukawa.kyoto-u.ac.jp/abs/q-bio.NC/0401009)}
} 
\end{center}

\begin{center}
Hideo Hasegawa
\footnote{Phone:+81-42-329-7482;
Fax: +81-42-329-7491;
E-mail:  hasegawa@u-gakugei.ac.jp}
\end{center}

\begin{center}
{\it Department of Physics, Tokyo Gakugei University  \\
Koganei, Tokyo 184-8501, Japan}
\end{center}
\begin{center}
(Received \today)
%{\rm (November 10, 2000)}
\end{center}
%\maketitle
\thispagestyle{myheadings}

\begin{center} 
{\bf Abstract}   \par
\end{center} 

\small

Effects of distractions such as
noises and parameter heterogeneity have been studied
on the firing activity of ensemble neurons, each of which
is described by the extended Morris-Lecar model showing 
the graded persisting firings with the aid of an included
${\rm Ca}^{2+}$-dependent cation current.
Although the sustained activity of {\it single} neurons 
is rather robust in a sense that the activity is realized
even in the presence of the distractions, 
the graded frequency of sustained firings is vulnerable to them.
It has been shown, however, that 
the graded persisting activity 
of {\it ensemble} neurons 
becomes much robust to the distractions
by the pooling (ensemble) effect. 
When the coupling is introduced,
the synchronization of firings in ensemble neurons is 
enhanced, which is beneficial to firings of target neurons.

\normalsize

\vspace{0.5cm}

\vspace{1.0cm}

\newpage
\section{Introduction}

It has been reported that persistent activity of neurons 
is ubiquitous in living brains.
Such persistent, stimulus-dependent activities are expected to
be neuronal substrates of the short-term working memory 
(Brody, Rome and Kepecs, 2003).
For example, in the prefrontal cortex of monkeys
which are trained by short-term tasks,
the persisting activity has been interpreted as a basis
of working memory for a few seconds 
(Funahashi, Bruce and Goldmanrakic, 1989;
Miller, Erickson and Desimone, 1996;
Romo, Hermandez, Lemus, Zainos and Brody, 2002).
In the area I of goldfish hindbrain which controls the motor systems
deriving the eye muscle, the persisting activity is interpreted
as a short-term memory of the eye position that keeps the eye still
between saccades 
(Pastor, Delacruz, and Baker, 1994;
Seung, 1996;
Aksay, Baker, Seung and Tank, 2000).  
In the later example, oculomotor neurons are considered to
integrate the velocity information from the eye movement,
yielding information on the eye position.

One of the problems of the observed persisting activity is 
how to keep it for extended periods.
Most models previously proposed are based on the recurrently
coupled neural networks 
(Rosen, 1972; 
CannonCRobinson and Shamma, 1983;
Seung, Lee, Reis and Tank, 2000, 
Miller, Brody, Romo and Eang, 2003).
These models may account for the observed property of 
the persistent activity. They have, however, a serious drawback 
requiring the fine tuning of the network-feedback
gain within a tolerance of less than 1\%\ to stabilize
the firing rate expressing stored information.
If the overall gain of the network is slightly increased
than the critical value, it leads to a divergent activity.
On the other hand, if the gain is slightly decreased than the critical value,
the state reduces to the stable fixed point without firings, losing
the memory stored in the firing rate. 
Some homeostatic mechanisms such as an activity-dependent
scaling of synaptic weights (Renart, Song and Wang, 2003) have been
proposed for stabilizing the neural activity.

A new mechanism based on the bistability of neurons
has been proposed 
(Rosen, 1972, Guigon, Dorizzi, Burnodand Schultz, 1995,
Koulakov, Raghavachari, Kepec and Lisman, 2002).
Two specific ways to realize bistable elements have
been discussed:
one by local excitation connections
and the other by the voltage-dependent
N-methyl-D-aspartate (NMDA) channel 
(Lisman, Fellous and Wang, 1998).
Recurrent networks consisting of 
bistable elements have been shown to robustly operate
even when the model parameters are not well tuned.
It is, however, unclear whether the prediction of the bistable
neuron is consistent with experimental data from
neuronal integrators.
For example, bistable neurons are reported to be absent 
in the goldfish oculomotor neurons
which show the persistent activity 
(Pastor {\it et al.}, 1994;
Aksay {\it et al.}, 2000).
%(Pastor, Delacruz, and Baker, 1994;
%Aksay, Baker, Seung and Tank, 2000).  

Quite recently, the graded persistent activity has been observed
for {\it single} neurons in {\it vitro} 
of the layer-V entorhinal cortex (EC) of 
rats 
(Egorov, Hamam, Fransein, Hasselmo and Alonso, 2002).
The sustained firing rate of the 
neurons is shown to be proportional to the integral over time of its
previous synaptic input.
EC in the parahippocampal region
is known to be associated with the working memory 
(Frank and Brown, 2003). 
It has been shown that EC neurons display
persisting activity during the delay phase of delayed mach 
or non-match to sample memory task (Suzuki, Muller and Desimone, 1997,
Young, Otto, Fox and Eichenbaum, 1997).
A phenomenological rate-code model including multiple
bistable dendrites has been proposed
(Goldman, Levine, Major, Tank and Seung, 2003).
It has been suggested that the calcium concentration is 
a plausible candidate for the observed parametric working memory
(Egorov, Hamam, Fransein, Hasselmo and Alonso, 2002).
Several conduction-based
neuron models have been proposed based on ${\rm Ca}^{2+}$-ion
channel.
Lowenstein and Sompolinksy (LS) (2003)
have shown that 
by the diffusion of the non-linear dynamics of 
the calcium concentration in bistable dendrites,
the wave-front of calcium propagates at a speed proportional to the 
synaptic inputs.
When the ${\rm Ca}^{2+}$-dependent cation is included in a neuron model
described by the HH-type model 
(Shriki, Hansel and Sompolinsky, 2003), calculated firings
show the behavior similar to the observed one.
Teramae and Fukai (2003) have proposed detailed
conductance-based models in which both L-type Ca channel
and ${\rm Ca}^{2+}$-dependent cation channel are taken into account.

It is interesting that in contrast to recurrent neural networks
with bistable neurons, single neurons with bistable dendrites
may store information in the form of the 
graded persisting activity.
For a reliable performance of information storage,
such single neurons are required to be robust against
distractions such as noises and heterogeneity in neuron elements.
It was reported (Mainen and Sejnowsky, 1995)
that although firings of single neurons 
in {\it vitro} are precise, those in {\it vivo} are quite
unreliable due to noisy environment.
This may suggest that noises in {\it vivo} might make 
the graded persisting activity
of EC neurons unreliable.
Experiments on EC neurons {\it in vivo} are needed
although they are difficult owing to the problem
in performing the intracellular recording in awake animals.

A small cluster of cortex generally consists of many similar neurons
which include some degree of heterogeneity.
Each neuron generates spikes which propagate to
synapses exciting neurons in the next stage.
It has been recognized that the population of neuron 
ensembles plays important roles in the information transmission.
The population improves the reliability of response of noisy neurons.
For example, the stochastic resonance in neuron ensembles
is much effective than that of single neurons.
An advantage of the synchronized activity is a large
input on a target neuron.

From these considerations, we will make, in this paper, 
a theoretical study on the robustness of the parametric 
working memory, 
taking account of two factors: (1) neuron ensembles
with (2) the heterogeneity and noises.
In order to investigate 
the property of the persisting activity
of an ensemble against noises and heterogeneity,
we will perform simulations on ensemble neurons,
each of which shows the persisting activity. 
We first develop a minimum, single neuron model, 
which is suitable for a simulation of ensemble neurons. 
The Morris-Lecar (ML) model, which was initially proposed for a
barnacle giant muscle fiber, has been
widely employed for a study on neuron dynamics
(Morris and Lecar, 1981; Rinzel and Ermentrout, 1989).
Although the ML model is the reduced, simplified model of 
the detailed HH model,
it is more realistic than the integrate-and-fire model.
A new variable for the calcium channel whose function
depends on ${\rm Ca}^{2+}$-ion concentration has been incorporated
to the ML model. 
With the aid of ${\rm Ca}^{2+}$-dependent current, 
a single ML neuron may show the graded persisting activity.

The paper is organized as follows. 
The property of a single (extended) ML model is discussed in Sec. 2,
where effects of noises and variations of model
parameters are investigated.
In Sec. 3, we have studied
a neuron ensemble where each neuron is described by the extended
ML model proposed in Sec. 2.
Effects of noises and
the heterogeneity of model parameters on
the persisting activity have been investigated.
The synchronization within neuron ensembles is also discussed.
Conclusions and discussions are given in Sec. V.

%\newpage

\section{Dynamics of a single neuron}
\subsection{Adopted model}

We have adopted the extended ML neuron model
for a single neuron,
given by
\begin{eqnarray}
C \frac{d v}{dt}
&=&-I_{Ca} - I_{K} -I_{cat} -I_{L}
+ a + I(t)+ \beta_v \:\xi(t), \\
\frac{d w_i}{dt}
&=& \phi \left[ \frac{w_{o}(v)-w}{\tau_{w}(v)} \right], \\
\frac{d z}{dt }&=& -\frac{z}{\tau_z}
+ b + d \:[I(t) + \beta_z \:\eta(t)], 
%\hspace{1cm} \mbox{($p=2$ to 4)}
\end{eqnarray}
where
\begin{eqnarray}
I_{Ca}&=& g_{\rm Ca} m_{o}(v)(v-v_{\rm Ca}), \\
I_{K}&=& g_{\rm K} \:w \:(v-v_{\rm K}),\\
I_{cat}&=& g_{\rm cat}\:z\:(v-v_{\rm cat}),\\
I_{L}&=& g_{\rm L} (v-v_{\rm L}), 
\end{eqnarray}
with
\begin{eqnarray}
m_{o}(v) &=& \frac{1}{2}
\left[ 1+tanh \left( \frac{v-v_1}{v_2} \right) \right], \\
w_{o}(v) &=& \frac{1}{2}
\left[ 1+tanh \left( \frac{v-v_3}{v_4} \right) \right], \\
\tau_{w}(v) &=& sech \left( \frac{v-v_3}{2 v_4} \right).
\end{eqnarray}
The model given by Eqs. (1)-(10) takes into account 
the non-specified cation current ($I_{cat}$)
besides Ca, K and leakage currents 
($I_{Ca}$, $I_{K}$ and $I_{L}$) which are included 
in the original ML model 
(Morris and Lecar, 1981; Rinzel and Ermentrout, 1989);
$v$ and $w$ denote the fast membrane potential
and the slow auxiliary variable, respectively;
$z$ expresses a new variable
for the ${\rm Ca}^{2+}$-ion concentration
which is relevant to the cation channel.
The variable $z$ controls the conductance of the cation channel  
which yields the persistent parametric activity,
as will be shown later.
The equation of motion for $z$
is assumed to be given by Eq. (3), where $b$, $d$ and  $\tau_z$
stand for the drift velocity, the coefficient, and  the life time,
respectively. 
Equation (3) may be justified,
to some extent, by the model proposed by LS
(Loewenstein and Sompolinsky, 2003), 
details being given 
in appendix A.
The explicit form of an input current $I(t)$ will
be shown shortly [Eq. (11)].
$\beta_v$ and $\beta_z$ express the strengths of 
independent white noises
given by $\xi(t)$ and $\eta(t)$ with zero means and
$<\xi(t)\:\xi(t')>=<\eta(t)\:\eta(t')>=\delta(t-t')$ and
$<\xi(t)\:\eta(t')>=0$.
In this study, we have adopted the following parameters:
the capacitance is $C=20$,
the reversal potentials of
Ca, K, cation channels and leakage are 
$v_{\rm Ca}=120$, $v_{\rm K}=-84$, $v_{cat}=40$
and $v_{\rm L}=-60$, and 
the corresponding conductances are
$g_{\rm Ca}= 4$,
$g_{\rm K}= 8$, $g_{cat}=1$ and
$g_{\rm L}= 2$:
other parameters are $v_1=-1.2$, $v_2=18$, $v_3=12$
$v_4=17.4$, $\phi=0.0667$ (Rinzel and Ermentrout, 1989), 
$a=39.6$, $b=0$, $d=0.0001$ and $\tau_z = \infty$:
values of $\beta_v$ and $\beta_z$ will be shown shortly
(in this paper, conductances are expressed in $mS/cm^2$,
currents in $\mu A/cm^2$, voltages in mV, the time in ms and 
the capacitance in $\mu {\rm F}/cm^2$).
These parameter have not necessarily been chosen so as
to reproduce observed data of EC (Egorov {\it et al.,} 2002).

\subsection{Calculated results}
\subsubsection{Property of the extended ML model}

It is noted that our ML model with these parameters 
can start firings from the zero frequency
when the parameter of $a$ is varied for $a_1 \leq a < a_2$ 
where the critical
values are $a_1=40$ and $a_2=116$ $\mu A/cm^2$ 
for $b=0$ and $I(t)=0$
[see Fig. 7.6 of Rinzel and Ermentrout, 1989].
Then our ML neuron belongs to the type-I neuron. 
Although the linearity of the $a-f$ relation at $a \geq a_1$ 
is not perfect, it does not matter for our purpose discussing effects
of distractions on the persisting activity.

We apply to our ML neuron,
an input signal consisting of four pulses as given by 
\begin{equation}
I(t)=\sum_{n=1}^4\: I_n(t)
= A_i \sum_{n=1}^{4} \;  
c_n [\Theta(t-T^{(i)}_{n})-\Theta(t-T^{(i)}_{n}-T_w)],
\end{equation}
where $A_i$ stands for the magnitude of a pulse,
$T^{(i)}_{n}= 1000\:n$ ms ($n$: integer), $T_w=200$ ms,
$c_n = 1$ ($-1$) for $n=1-3$ ($n=4$), 
and $\Theta(t)$ is the Heaviside function.
We have adopted $A_i=20$ $\mu A/cm^2$ otherwise noticed.

Equations (1)-(11) have been solved by the fourth-order Runge-Kutta
method with a time step of 0.01 ms, the initial condition being given by
$v(0)=-40$, $w(0)=0$ and $z(0)=0$.
Figure 1(a) shows time courses of $v$, $w$, and $z$
when $I(t)$ given by Eq. (11) is applied to our ML neuron,
$I(t)$ being depicted at the bottom of Fig. 1(a).
The variable $z$ is proportional to the integral on time of
an input signal [Eq. (3)], and it shows a step-wise behavior:
$z$ is changed while an input pulse given by Eq. (11) is added
and it stays at plateau values of $z=0.02$, 0.04 and 0.06 
between input pulses.
Figure 1(b) shows time courses of currents of $I_{K}$, $I_{Ca}$
and $I_{cat}$. It is noted that $I_{Ca}$ and $I_{cat}$ are
inward (negative) currents while $I_K$ is the outward (positive) one.
When an input pulse is applied to a ML neuron, $z$ is increased,
which triggers a flow of $I_{cat}$ and the oscillation of state
variables. As $z$ is furthermore increased by an injected pulse,
the frequency of firing is increased. 
This is more clearly seen by circles
in Fig. 1(c), which shows the time-dependent frequency $f(t)$ defined as
the inverse of the interspike interval (ISI) given by
\begin{equation}
T_{n}(t_n)=t_{n+1}-t_n,
\end{equation}
$t_n$ being the $n$th firing time defined as the time
when $v(t)$ crosses the threshold $\theta$ ($-10$ mV) from below.
Neurons fire most strongly during the transient response
to applied pulses and settle to a persistent firings
which depend on the integral on time of the past inputs. 
An initial 200-ms depolarizing pulse results in a sustained firing of 5.1 Hz,
and the second and third ones increase the frequency to 7.6 and 9.0 Hz,
respectively, while a persisting firing backs to
7.6 Hz after the fourth hyperpolarizing pulse.  
We note that the firing frequency $f(t)$ is successively increased 
(decreased) when depolarizing (hyperpolarizing) pulses are applied,
and that the frequency is constant with plateaus between applied pulses.
This trend is realized also in
histogram in Fig. 1(d) expressing the firing rate $r(t)$ for
various time bins of $T_b=200$ (solid curve), 
400 (dashed curve) and 600 (chain curve). 
The rate with the small time bin of $T_b=200$ reproduces
the transient behavior while impulses are applied, although it
shows an oscillation at the plateau period.
In contrast, the rate with large $T_b=600$, the persisting 
firings are well explained while the transient firings are
averaged out.
When the magnitude of pulses 
$A_i$ is increased (decreased), the value of $f(t)$ 
at plateau is increased (decreased),
as shown by triangles ($A_i=30$) and 
inverted triangles ($A_i=10$) in Fig. 1(c). For example,
the first depolarizing pulse with $A_i=$ 10, 20 and 30 $\mu A/cm^2$
yields the sustained firing frequencies of 2.9, 5.1 and 6.5 Hz, respectively.
Similarly, when the duration of pulses $T_w$ is increased (decreased),
the value of $f(t)$ at plateau is increased (decreased)
(results are not shown).
Calculated results are similar to those experimentally obtained for single
EC neurons. 
(Egorov {\it et al.,} 2002).

Figures 2(a) and 2(b) show the $v-w$ and $z-v$ phase planes
for firings shown in Fig. 1.
For a comparison, nullclines of Eqs. (1) and (2) for $a=39.6$,
$b=0$, $z=0$ and $I=0$ are shown by dashed and chain curves,
respectively,
which have a tangentially contact at $(v,w)=(-29.6, \:0.00846)$.
The solid curve in Fig. 2(a) shows that the cycle starts from 
$(v,w)=(-40,\:0)$ and circles counter-clockwise
almost independently of the $z$ value.
Figure 2(b) shows that between the plateau values of 
$z=0.02$, 0.04 and 0.06, $v$ oscillates against $z$, 
and that $v$ changes vertically when $z$ stays at these plateau values.  

\subsubsection{Effects of model parameters of $a$ and $b$}

We will investigate effects of parameters when their 
values are changed.
Among many model parameters included in Eqs. (1)-(10), 
we have chosen $a$ and $b$ as the parameters to be changed
because they are expected to play an important role
in stabilizing the persisting firings.
Figure 3(a) shows the time course of $f(t)$ for various $a$ values of
$a=$39, 39.6 and 41 $\mu A /cm^2$ with $b=0$.
For $a=41$, a neuron fires with $f=5.2$ Hz at $t < 1000$ ms
before the first pulse is applied, and
firing frequencies after the second, third and fourth inputs
are larger than those for $a=39.6$.
In contrast, the firing frequency for $a=39$
is smaller than that for $a=39.6$.

Figure 3(b) shows $f(t)$ for various $b$ values
of $b=-2$, 0, and 2 ($\times 10^{-6}$) with $a=39.6$ $\mu A/cm^2$.
The result for $b=0$ is the same as that shown by circles in Fig. 1(c).
For $b=-2 \times 10 ^{-6}$, $f(t)$ between pulses is gradually decreased. 
For $b=2 \times 10 ^{-6}$, in contrast,
the frequency shows a positive drift.

\subsubsection{Effects of white noises}

Next we add white noises to our ML neuron.
The time course of the frequency $f(t)$ when white noises 
of $\beta_v=4$ are added
to the variable $v$ in Eq. (1), is shown by dots in Fig. 3(c),
where open circles express the result with no noises.
Because of added noises, the frequency of persisting
firings fluctuates.
A similar behavior has been realized
when noises are added to the variable $z$ in Eq. (3), 
whose result for $\beta_z=$ 2 is shown by dots in Fig. 3(d).

Figures 3(a)-3(d) show that a single neuron is rather robust 
in a sense that
the persisting firings are possible even when distraction are added.
We note, however, that the frequency of the persisting firings 
is modified by distractions.
The robustness is much improved 
in ensemble neurons by the pooling effect,
as will be discussed in the following Se. 3.

\section{Dynamics of neuron ensembles}
\subsection{Adopted model}
We have assumed that an ensemble consisting
of $N$-unit ML neurons stores information
from an applied signal of $I(t)$.
Dynamics of an adopted the extended ML neuron model is given by
\begin{eqnarray}
C \frac{d v_i}{dt}
&=& g_{\rm Ca} m_{o}(v_i)(v_{\rm Ca}-v_i)
+ g_{\rm K} w_i (v_{\rm K}-v_i) 
+ g_{\rm cat}\:z_i\:(v_{\rm cat}-v_i) \\
&&+ g_{\rm L} (v_{\rm L}-v_i)
+ a_i + I_i(t)+\beta_v \:\xi_i(t), \\
\frac{d w_i}{dt}
&=& \phi \left[ \frac{w_{o}(v_i)-w_i}{\tau_{w}(v_i)} \right], \\
\frac{d z_i}{dt }&=& -\frac{z_i}{\tau_z}
+ b_i + d \:[I_i(t)+\beta_z \:\eta_i(t)], \\
I_i(t)&=& I(t)
+ \left( \frac{J}{N-1} \right) \sum_{j (\neq i)} G(v_j),
\hspace{1cm} \mbox{($i=1$ to $N$)}
\end{eqnarray}
where the sigmoidal function $G(v)$ 
is given by $G(v)=1/[1+exp(v-\theta)/\alpha]$ 
with the coupling constant $J$,
the threshold $\theta$ ($=-10$ mV) and the width $\alpha$ ($=1$),
self-coupling terms being excluded.
$\beta_v$ and $\beta_z$ express the
strengths of independent white noises of $\xi_i(t)$ and $\eta_i(t)$
with zero means and
$<\xi_i(t)\:\xi_j(t')>= <\eta_i(t)\:\eta_j(t')>
=\delta_{ij} \delta(t-t')$ and
$<\xi_i(t)\:\eta_j(t')>=0$;
other notations are the same as in Eqs.(1)-(10).
The model parameters of $a_i$ and $b_i$ are assumed to
be heterogenous and their distributions obey the Gaussian distribution  
with
\begin{eqnarray}
<a_i> &=& \epsilon, \\
<\delta a_i \: \delta a_j> &=& \lambda^2 \:\delta_{ij},\\
<b_i> &=& 0, \\
<b_i \: b_j> &=& \nu^2 \:\delta_{ij},
\end{eqnarray}
where $\delta a_i=a_i-\epsilon$.
We have adopted the same model parameters as for a single
ML model discussed in Sec. 2.1 otherwise noticed.

\subsection{Calculated results}
\subsubsection{Homogeneous ensembles with no noises}

When we perform simulations for 10-unit homogeneous
ML neuron with no noises and no couplings
($\epsilon=39.6$, $\lambda=0$, $\nu=0$,
$\beta_v=0$ and $\beta_z=0$, and $J=0$), we get
the result in which time courses of state variables of $v$, $w$ and $z$
of all neurons in the ensemble
are the same as that shown in Fig. 1(a) for a single ML neuron;
$f(t)$ and $r(t)$ are also the same as those in Fig. 1(c) and 1(d).
In the following, we will investigate effects of the heterogeneity 
and noises on the persisting activity in ML neuron ensembles.

\subsubsection{Effects of heterogeneity}

Raster in Fig. 4(a) shows firings of neuron ensembles
when the heterogeneity of $\lambda=1$ is introduced 
to the parameter of $a_i$. Raster shows that
due to the introduced heterogeneity, firings of some neurons
are increased while those of some neurons are decreased,
compared to those in homogeneous neuron ensembles. 
For example, five neurons with large $a_i$ fire
before the first impulse is applied, 
which yields a finite firing rate at $t < 1000$ ms.
Nevertheless, the firing rate averaged
over the ensemble is not so different from that
of the homogeneous ensemble shown in Fig. 1(d). 

Raster in Fig. 4(b) shows firings of an ensemble
when the heterogeneity of $\nu=2 \:\times 10^{-6}$ 
is introduced to the parameter of $b_i$  .
The firing becomes irregular due to the introduced heterogeneity.
However, the histogram of firing rate averaged over the ensemble
shows a stable, persisting firing. 

\subsubsection{Effects of noises}

We have added white noises to a ML neuron ensemble.
Raster of Fig. 4(c) shows firing of the ensemble when noises with
$\beta_v=4$ are added to the variable $v$.
Because of added noises,
firings of an ensemble become irregular. 
Despite added noises, a stable, persisting 
firing is realized as the averaged firing rate $r(t)$ shows,

When white noises of $\beta_z=2$ is added to the variable $z$,
firings of an neuron ensemble become irregular, as 
raster in Fig. 4(d) shows. We note from the histogram that
firings show a stable, persisting activity.

\subsubsection{Effects of couplings}

We have so far neglected the couplings among neurons ($J=0$)
which are now taken into account. 
In order to examine the firing synchrony
in an ensemble, we consider the quantity given by
\begin{equation} 
R(t)= \frac{1}{N^2} \sum_i \sum_j <[v_i(t)-v_j(t)]^2>,
\end{equation}
which is zero in the completely synchronized state.
By using a proper normalization factor of
$R_0(t)=2 (1-1/N) \gamma(t)$ which
expresses the $R(t)$ value for an asynchronous state, we
define the {\it synchronization ratio} given by 
(Hasegawa, 2003a, 2003b)
\begin{equation}
S(t)=1-\frac{R(t)}{R_0(t)} = \frac{\zeta(t)}{\gamma(t)},
\end{equation} 
where
\begin{eqnarray}
\zeta(t)&=& \frac{1}{N(N-1)} \sum_i \sum_{j (\neq i)} 
<\delta v_i(t) \: \delta v_j(t)>, \\
\gamma(t)&=& \frac{1}{N} \sum_i <\delta v_i(t)^2>,
\end{eqnarray}
with $\delta v_i(t)=v_i(t)-<v_i(t)>$.
It is easy to see that $S(t)$ is 1 and 0 
for the completely synchronous and asynchronous states,
respectively.

The lower frame of Figs. 5(a) shows
the time course of the frequency $f(t)$ calculated for an ensemble 
with the heterogeneity of $\lambda=1$ 
and no couplings ($J=0$) by 100 trials.  This should be compared to
the histogram for the rate shown in Fig. 4(a),
which has been calculated by a single trial.
The middle frame of Fig. 5(a) expresses the synchronization 
ratio $S(t)$, which is vanishing because
of the heterogeneity (with no couplings).
Lower frames of Figs. 5(b), 5(c) and 5(d) show $f(t)$
for introduced couplings of $J=20$, 40 and 60, respectively.
The distribution of $f(t)$ for $J=20$ become wider than that
for $J=0$. This may be understood as follows.
When any neuron with a large $a_i$ fires, its spike propagates
through couplings and induces firings in other neurons
of an ensemble. When the coupling is furthermore increased, 
however, the distribution in $f(t)$ is reduced 
as the lower frame of Fig. 5(d) for $J=60$ shows.
In contrast, the synchronization ratio $S(t)$ shown in the middle frames
of Fig. 5(b), (c) and (d) is gradually increased as $J$ is increased.
It is interesting to note that the synchronization ratio is
enhanced by transient impulses, and that $S(t)$ is gradually developed
as more impulses are applied. 

Figures 6(a)-6(d) show similar plots of $f(t)$ and $S(t)$ 
for ensembles with various couplings
subject to noises of $\beta_v=4$, calculated by 100 trials.
Lower and middle frames of Fig. 6(a) show that for $J=0$,
$f(t)$ has a wide distribution and $S(t)$ is zero
due to noises. 
When the coupling is introduced, the distribution of $f(t)$
is once increased for $J=20$, but reduced for stronger
$J=40$ and 60 while the synchronization ratio $S(t)$ 
is monotonously increased as increasing $J$. 
The $J$ dependence of $f(t)$ and $S(t)$ in Figs. 6(a)-6(d)
is the same as that in Figs. 5(a)-5(d).

%\newpage
\section{Conclusion and Discussion}

In this paper, we have first proposed a minimum,
conductance-based model
showing the graded persisting activity
by incorporating a calcium channel (Lowenstein and Sompolinsky, 2003)
to the ML model (Morrris and Lecar, 1981; Rinzel and Ermentrout, 1989).
Then, by employing the extended ML model, 
we have studied effects of distractions such as 
noises and the heterogeneity in model parameters
on the persisting activity of single and ensemble neurons.
The activity of single ML neurons
is vulnerable because the frequency 
of persisting firings is modified by the distractions.
In particular, even a small $b$ ($=2 \times 10^{-6}$)
yields a slow drift in the sustained frequency [Fig. 3(b)].
This is partly due to the fact that the value of the variable $z$
is very small ($\sim 0-0.06$) compared to those of $v$ and $w$.
As discussed in appendix A, a vanishing of the drift requires $b=0$,
which is realized when parameters of $c_1$, $c_2$ and $c_3$
satisfy the condition given by $c_2=(c_1+c_3)/2$ [Eq. (A13)].
This condition may be relaxed when a given neuron has multidendritic
branches, for which $b$ is expressed by $b=<b_m>_m$, $b_m$ being
the $b$ term relevant to the $m$th dendrite and $<>_m$ the average
over multidendrites.
The averaging over multidendrites is expected to yield $b=0$ for 
single neurons.
Even if $b$ remains finite (but small), the drift of the sustained 
frequency may vanish in neuron ensembles [Fig. 4(b)]
because of the pooling (ensemble) effect which is widely observed in
various neuronal functions.
Although we have made simulations only for several typical sets of
parameter values,
they have shown that the graded persisting activity
of a neuron ensemble becomes more robust against distractions
than that of a single neuron again by the pooling effect

The temporally correlated neuronal activity,
the synchrony, has been considered to be important to various neuronal
processing such as perception 
(Gray and Singer 1989, Gray, K\"{o}nig, Engel and Singer, 1989)
and attention (Steinmetz et al. 2000,
Fries, Reynolds, Rorie and Desimone 2001).
Recently the temporal structure in neuronal activity
during working memory has been observed in parietal 
cortex of monkey (Pesaran, Pezaris, Sahani, Mitra and Andersen, 2002).
Our calculations have shown that 
couplings among neurons enhance the synchrony, which is
expected to be beneficial to firings of target neurons.
It is necessary to make more detailed study
on the interaction and synchrony in neuron
ensembles with the persisting activity.
Our calculations in this study have been based on direct
simulations. We are under consideration to
apply, to the ML model, a semi-analytical dynamical
mean-field theory which was successfully applied to large-scale
neuron ensembles described by FitzHugh-Nagumo 
and Hodgkin-Huxley models (Hasegawa, 2003a, 2003b).

\section*{Acknowledgements}
This work is partly supported by
a Grant-in-Aid for Scientific Research from the Japanese 
Ministry of Education, Culture, Sports, Science and Technology. 

\newpage

\noindent
{\large \bf Appendix A.  Derivation of Eq.(3)}

Assuming a long, linear dendrite for $x \in [-L, \:L]$,
Lowenstein and Sompolinsky (LS) (2003) have shown that 
the ${\rm Ca}^{2+}$-dependent cation current is
given by
\begin{eqnarray}
I_{cat}(t)&=&g_{cat}\: z(t) \:(v-v_{cat}), \nonumber 
\hspace{4cm}\mbox{(A1)}
\end{eqnarray}
with the spatially-summed ${\rm Ca}^{2+}$ concentration of $z(t)$
given by
\begin{eqnarray}
z(t)&=& \int_{-L}^{L} dx \; c(x,t), 
\nonumber\hspace{5cm}\mbox{(A2)}
\end{eqnarray}
where $c(z,t)$ stands for the space- and time-dependent 
${\rm Ca}^{2+}$ concentration satisfying the 
partial differential equation given by
\begin{eqnarray}
\frac{\partial c(x,t)}{\partial t}
= h(c(x,t)) + D \frac{\partial^2 c(x,t)}{\partial x^2}
+ k(c(x,t)) \:I(t).
\nonumber\hspace{2cm}\mbox{(A3)}
\end{eqnarray}
Here $h(c)=-K(c-c_1)(c-_2)(c-c_3)$
and $k(c)=(K/2)(c_3-c_1)(c-c_1)(c-c_3)$
with $c_1 < c_2 < c_3$, $I(t)$ denotes an input signal,
$D$ the diffusion constant and $K$ the positive coefficient.
The functional form of $h(c)$ shows that
the state is bistable at $c=c_1$ and $c=c_3$
but unstable at $c=c_2$.
LS have shown that
the solution of Eq. (A3) for the boundary condition
of $c(-L,t)=u_3$ and $c(L,t)=u_1$,
is given by
\begin{eqnarray}
c(x,t)=c_2+ \left( \frac{c_3-c_1}{2} \right) 
\:tanh \left[ \frac{x-P(t)}{\lambda} \right],
\nonumber\hspace{2cm}\mbox{(A4)}
\end{eqnarray}
with 
\begin{eqnarray}
P(t)=P(0)+u t +s \int_0^t \:dt' \;I(t'),
\nonumber\hspace{2cm}\mbox{(A5)}
\end{eqnarray}
where 
\begin{eqnarray}
\lambda&=&\left( \frac{2}{c_3-c_1} \right)
\sqrt{\frac{2D}{K}}, \nonumber 
\hspace{3cm}\mbox{(A6)}\\
u &=& \sqrt{2DK} 
\left[c_2-\left( \frac{c_1+c_3}{2} \right)\right], \nonumber 
\hspace{2cm}\mbox{(A7)}\\
s&=&(c_3-c_1) \sqrt{2 D K}. 
\nonumber\hspace{3cm}\mbox{(A8)}
\end{eqnarray}
Equations (A4)-(A8) imply
that the position of the front
of ${\rm Ca}^{2+}$-concentration moves with the velocity
of $u$, and it is proportional to
the integral of the input signal $I(t)$.
By using Eqs. (A1)-(A8), we get
\begin{eqnarray}
\frac{d z}{dt} 
&=& \left( \frac{c_3-c_1}{2} \right) 
\left( tanh\left[\frac{-L-P(t)}{\lambda}\right]
-tanh \left[ \frac{L-P(t)}{\lambda} \right] \right) \; \frac{dP(t)}{dt},
\nonumber\hspace{2cm}\mbox{(A9)} \\
&\simeq & b + d \:I(t),
\nonumber\hspace{10cm}\mbox{(A10)}
\end{eqnarray}
with
\begin{eqnarray}
b &=& (c_3-c_1) \:u, 
\nonumber\hspace{2cm}\mbox{(A11)}\\
d &=& (c_3-c_1) \:s,
\nonumber\hspace{2cm}\mbox{(A12)}
\end{eqnarray}
leading to Eq. (3), where the first term is included
for the relaxation process of $z$.
Equations (A8) and (A11) show that $b$ does not vanish unless parameters
of $c_1$, $c_2$ and $c_3$ satisfy the condition:
\begin{eqnarray}
c_2=\frac{c_1+c_3}{2},
\nonumber\hspace{2cm}\mbox{(A13)}
\end{eqnarray}
although LS adopted $b=0$ assuming the condition given by Eq. (A13).
%we have examined the effect of $b$ in our study.

For a single neuron
with $M$-unit multiple dendritic branches where the same signal is applied
to all synapses, the equation of motion for the averaged variable of 
$z=\sum_{m=1}^{M} z_m \equiv <z_m>_m$ 
is again given by Eq. (A10) but with
$b = <b_m>_m$ and $d = <d_m>_m$ where
$z_m$, $b_m$ and $d_m$ denote quantities relevant
to the $m$th dendrite.
%The other advantage of multidendrite computation is to
%reduce the effect of noises
%as pointed out by LS (Loewenstein and Sompolinsky, 2003).

\newpage
\noindent
{\large \bf References}

\begin{description}
%\begin{references}

\item Aksay E., Baker R., Seung H. S., and Tank D. W. (2000).
Anatomy and discharge properties of pre-motor neurons in the 
Goldfish medulla that have eye-position signals during fixations.
{\it Journal of Neurophysiology,} {\bf 84}, 1035-1049.

\item Brody C. D., Rome R., and Kepecs A. (2003).
Basic mechanisms for graded persistent activity:
discrete attractors, continuous attractors, and
dynamic representations.
{\it Current Opinion on Neurobiology,} {\bf 13}, 204-211; 
related references therein.

\item Camperi M. and Wang X. J. (1998).
A model of visuospatial working memory in prefrontal cortex: 
recurrent network and cellular bistability.
{\it Journal of Computational Neuroscience,} {\bf 5}, 383-405.

\item Cannon S. C.CRobinson D. A.. and Shamma  S.(1983).
A proposed neural network for the integrator of the oculomotor systems.
{\it Biological Cybernetics,} {\bf 49}, 127-136.

\item Egorov A. V., Hamam B. N., Fransein E.,
Hasselmo M. E., and Alonso A. A. (2002).
Graded persistent activity in entorhinal cortex neurons.
{\it Nature,} {\bf 420}, 173-178. 

\item Frank L. M. and Brown E. N. (2003).
Persistent activity and memory in the entorhinal cortex.
{\it Trends in Neurosciences,} {\bf 26}, 400-401.

\item Fries P., Reynolds J. H., Rorie A. H. and Desimone R., (2001).
Modulation of oscillatory neuronal synchronization by
selective visual attention.
{\it Science,} {\bf 291}, 1560-1563.

\item Funahashi S., Bruce C. J., and Goldmanrakic P. S. (1989).
Mneumonic coding of visual space in the monkeys dorsolateral cortex.
{\it Journal of Neurophysiology,} {\bf 61}, 331-349.

\item Goldman M. S., Levine J. H., Major G., Tank D. W.
and Seung H. S. (2003).
Robust persistent neural activity in a model integrator with
multiple hysteretic dendrites per neuron. (preprint) 

\item Gray C. M., K\"{o}nig P., Engel A. K. and Singer W. (1989).
Oscillatory responses in cat visual cortex exhibit inter-columnar
synchronization which reflects global stimulus properties.
{\it Nature,} {\bf 338}, 334-337.

\item Gray C. M. and Singer W. (1989).
Stimulus-specific neuronal oscillations in orientation 
columns of cat visual cortex.
{\it Proceedings of National Academy of Sciences
of the United State of America,} {\bf 86}, 1698-1702.

\item Guigon E., Dorizzi B., Burnod Y. and Schultz W. (1995).
Neural correlates of learning in the prefrontal cortex
of monkey.
{\it Cerebrum Cortex,} {\bf 2}, 135-147.

\item Hasegawa H., (2003a)
Dynamical mean-field theory of spiking neuron ensembles:
Response to a spike with independent noises. 
{\it Physical Review E,} {\bf 67}, 041903.1\\-041903.19.

\item Hasegawa H., (2003b)
Dynamical mean-field theory of noisy spiking neuron ensembles:
Application to Hodgkin-Huxley model. 
{\it Physical Review E,} {\bf 68} 041909.1-041909.13.

\item Koulakov A. A., Raghavachari S., Kepec A., and Lisman J. E. (2002).
Model for a robust neural integrator.
{\it Nature Neuroscience,} {\bf 5}, 775-782.

\item Lisman J. E., Fellous J. M., and Wang X. J. (1998).
A role for NMDA-receptor channels in working memory.
{\it Nature Neuroscience,} {\bf 1}, 273-275.

\item Loewenstein Y. and H. Sompolinsky (2003).
Temporal integration by calcium dynamics in a model neuron.
{\it Nature Neuroscience,} {\bf 6}, 961-967

\item Mainen Z. F., and Sejnowsky T. J. (1995).
Reliability of spike timing in neocortical neurons.
%rat cortical slice
{\it Science,} {\bf 268}, 1503-1506.

\item Miller E. K., Erickson C. A., and Desimone  R. (1996).
Neural mechanisms of visual working memory in prefrontal cortex 
of macaque.
{\it The Journal of Neuroscience,} {\bf 16}, 5154-5167.

\item Miller P., Brody C. D., Romo R. and Wang X. (2003).
A recurrent network model of somatosensory parametric working
memory in the prefrontal cortex,
{\it Cerebrum Cortex,} {\bf 13}, 12080-1218

\item Morris C. and Lecar H. (1981).
Voltage oscillations in the barnacle giant muscle fiber.
{\it Biophysics} {\bf 35}, 193-213.
  
\item Pastor A. M., Delacruz R. R., and Baker R. (1994).
Eye position and eye velocity integrators reside in separate brain-stem nuclei.
{\it Proceedings of National Academy of Sciences
of the United State of America,} {\bf 91}, 807-811. 

\item Pesaran B., Pezaris J. S., Sahani M., Mitra P. P.
and Andersen R. A. (2002).
Temporal structure in neuronal activity during working memory
in macaque parietal cortex.
{\it Nature Neuroscience,} {\bf 5}, 805-811.

\item Renart A., Song P. and Wang X. (2003).
Robust spatial working memory through homeostatic synaptic
scaling in heterogeneous cortical networks.
{\it Neuron,} {\bf 38}, 473-485.

\item Rinzel J. R. and Ermentrout G. (1989).
Analysis of neural excitability and oscillation.
In C. Koch and I. Segev, (Eds.), 
{\it Methods in neural modeling} (pp251-291),
MIT press, Cambridge, MA.

\item Romo R., Hermandez A., Lemus L., Zainos A., and Brody C. D. (2002).
Neuronal correlates of decision-making in secondary 
somatosensory cortex.
{\it Nature Neuroscience,} {\bf 5}, 1217-1225. 

%recurent network
\item Rosen M. (1972). 
A theoretical neural integrator.
{\it IEEE Transactions on Biomedical Engineering,} {\bf 19}, 362-367.

\item Seung H. S. (1996).
How the brain keeps the eyes still.
{\it Proceedings of  National Academy of Sciences
of the United State of America,} {\bf 93}, 13339-13344.

\item Seung H. S., Lee D. D., Reis B. Y., and Tank D. W. (2000).
The autapse: A simple illustration of short-term analog memory
storage by tuned synaptic feedback.
{\it Journal of Computational Neuroscience,} {\bf 9}, 171-185. 

\item Shriki O., Hansel D., and Sompolinsky H. (2003).
Rate models for conductance-based cortical
neuronal networks.
{\it Neural Computation,} {\bf 15}, 1809-1841.

\item Steinmetz P. N., Roy A., Fitzgerald P. J.,
Hsiao S. S., Johnson K. O. and Niebur E. (2000).
Attention modulates synchronized neuronal firing in primate 
somatosensory cortex.
{\it Nature,} {\bf 404}, 187-190.

\item Suzuki W. A., Miller E. K. and Desimone R. (1997).
Object and place memory in the macaque entorhinal cortex.
{\it Journal of Neurophysiology,} {\bf 78}, 1062-1081.

\item Taramae J. and Fukai T. (2003), private communications.

\item Young B. J., Otto T., Fox G. D. and Eichenbaum H. (1997).
Memory representation within parahippocampal region.
{\it The Journal of Neuroscience} {\bf 17}, 5183-5195.

%\end{references}
\end{description}

%\noindent
%Goldman M. S., Levine J. H., Major G., 
%D. W. Tank, and H. S. Seung,

%\end{thebibliography}
\newpage

\begin{figure}
\caption{
%Fig B. 
Time courses of 
(a) $v$, $w$, $z$ and $I$ and
(b) $I_K$, $I_{cat}$ and $I_{Ca}$
of a single ML neuron with 
$A_i=20$, $a=39.6$ and $b=0$;
$w$, $z$ and $I$ are multiplied by factors of 100, 500,
and 10, respectively,
and are shifted downward by 100, 150, and 180, respectively;
$I_{cat}$ is multiplied by a factor of 100, and
$I_{Ca}$ is shifted downward by 30.
(c) The time course of the frequency $f(t)$ which is the inverse of
ISI, with $A_i=10$ (inverted triangles), 20 (circles)
and 30 (triangles).
(d) The histogram depicting the rate $r(t)$ of firings
shown in (a) and (b) with the time bins of 
$T_b=$ 200 (the solid curve), 400 (the dashed curve) 
and 600 (the chain curve).
}
\label{fig1}
\end{figure}

\begin{figure}
\caption{
%Fig A. 
(a) The $v-w$ and (b) $z-v$ phase planes
relevant to firings shown in Fig. 1;
dashed and chain curves in (a) express nullclines
of Eqs. (1) and (2)
for $A_i=20$, $a=39.6$, $b=0$, $z=0$ and $I=0$.
}
\label{fig2}
\end{figure}

\begin{figure}
\caption{
%Fig C. 
Time courses of the frequency $f(t)$ of a single ML neuron
(a) for $a=41$ (triangles), 39.6 (circles)
and 39 (inverted triangles) with $b=0$, $\beta_v=0$ and $\beta_z=0$,
(b) for $b=2 \times 10^{-6}$, 0 (circles)
and $-2 \times 10^{-6}$ (inverted triangles)
with $a=39.6$, $\beta_v=0$ and $\beta_z=0$,
(c) for $a=39.6$, $b=0$, $\beta_v=4$, and $\beta_z=0$, and 
(d) for $a=39.6$, $b=0$, $\beta_v=0$, and $\beta_z=2$:
an input signal is shown at bottoms, and
results of (c) and (d) are calculated by ten trials,
}
\label{fig3}
\end{figure}

\begin{figure}
\caption{
%Fig. E. 
Rasters showing firings of 10-unit
ML neuron ensembles and 
histograms expressing the firing rate $r(t)$
with time bins of $T_b=$ 200 (solid curves),
400 (dashed curves) and 600 (dot-dashed curves) 
for 
(a) $\lambda=1$, $\nu=0$, $\beta_v=0$, $\beta_z=0$,
(b) $\lambda=0$, $\nu=2 \times 10^{-6}$, $\beta_v=0$, $\beta_z=0$,
(c) $\lambda=0$, $\nu=0$, $\beta_v=4$, $\beta_z=0$ and
(d) $\lambda=0$, $\nu=0$, $\beta_v=0$ $\beta_z=2$,
with $\epsilon=39.6$ and $J=0$, calculated by a single trial,
an input signal being shown at bottoms. 
}
\label{fig4}
\end{figure}

\begin{figure}
\caption{
%Fig. K. 
Time courses of the frequency $f(t)$ (lower frames)
and the synchronization ratio $S(t)$ (middle frames) 
of heterogeneous ML neuron ensembles for 
(a) $J=0$, (b) $J=20$, (c) $J=40$ and (d) $J=60$,
with $\lambda=1$, $\beta_v=0$,
$\epsilon=39.6$ and $\beta_z=0$,
calculated by 100 trials; an input signal is shown in upper frames, and 
left and right ordinates are for $f(t)$ and $S(t)$, respectively.
}
\label{fig5}
\end{figure}

\begin{figure}
\caption{
%Fig. K. 
Time courses of the frequency $f(t)$ (lower frames)
and the synchronization ratio $S(t)$ (middle frames) 
of noisy ML neuron ensembles for 
(a) $J=0$, (b) $J=20$, (c) $J=40$ and (d) $J=60$,
with $\beta_v=4$, $\lambda=0$, 
$\epsilon=39.6$ and $\beta_z=0$,
calculated by 100 trials; an input signal is shown in upper frames, and 
left and right ordinates are for $f(t)$ and $S(t)$, respectively.
}
\label{fig6}
\end{figure}

%------------------------------------
\end{document}